# An achiral magnetic photonic antenna as a tunable nanosource of superchiral light

Lingfei Cui[1], Xingyu Yang[1], Benoît Reynier[1], Catherine Schwob[1], Sébastien Bidault[2], Bruno Gallas[1,]* and Mathieu Mivelle[1,]*

[1]Sorbonne Université, CNRS, Institut des NanoSciences de Paris, INSP, F-75005 Paris, France

[2]Institut Langevin, ESPCI Paris, PSL University, F-75005 Paris, France

*Corresponding authors:

bruno.gallas@insp.jussieu.fr

mathieu.mivelle@sorbonne-universite.fr

ORCID: 0000-0002-0648-7134

## Abstract

Sensitivity to molecular chirality is crucial for many fields, from biology and chemistry to the pharmaceutical industry. By generating superchiral light, nanophotonics has brought innovative solutions to reduce the detection volume and increase sensitivity at the cost of a non-selectivity of light chirality or a strong contribution to the background. Here, we theoretically propose an achiral plasmonic resonator, based on a rectangular nanoslit in a thin gold layer behaving as a magnetic dipole, to generate a tunable nanosource of purely superchiral light. This nanosource is free of any background, and the sign of its chirality is externally tunable in wavelength and polarization. These properties result from the coupling between the incident wave and the magnetic dipolar character of our nano-antenna. Thus, our results propose a platform with deep subwavelength detection volumes for chiral molecules in particular, in the visible, and a roadmap for optimizing the signal-to-noise ratios in circular dichroism measurements to reach single-molecule sensitivity.

## INTRODUCTION

Molecular chirality, the lack of superposition under mirror reflection, is of significant importance in biological systems as well as in biological and chemical processes. For instance, biological taste and smell receptors are sensitive to enantiomers, the two reflected images of a chiral molecule, and can differentiate them chemically by producing different responses that we interpret, for example, as radically different smells.[1] This asymmetry is essential in the case of physiological drug action, where, in the worst case, one enantiomer acts as a drug while the other has harmful effects.[2] Chiral molecules reveal their chirality through interaction with another chiral entity, such as circularly polarized light. Circular dichroism (CD) spectroscopy exploits the difference in light absorption for right and left circular polarizations to detect and identify chiral molecules. This is why intense research based on optical nanostructures has been developed for several years, aiming at reducing the detection volume while increasing the sensitivity. This research is based on the concept of superchiral light, defined as a chirality density higher than propagative circularly polarized light.[3] Indeed, the CD is proportional to the chirality density $C = -\frac{\varepsilon_0 \omega}{2} Im(\boldsymbol{E}^* . \boldsymbol{B})$, where $\varepsilon_0$ and $\omega$ are the permittivity of vacuum and the angular frequency, respectively, and **E** and **B** are the time dependent electric and magnetic fields.[4-6] Although the interaction between a chiral medium and an optical nanostructure is a complex mechanism,[7-10] the ability of resonant optical nanostructures, such as photonic nano-antennas, to enhance the electric and magnetic optical near-fields allows to maximize C, and thus the CD of molecules in their vicinity, increasing the sensitivity in enantioselectivity while decreasing the interaction volume.

So far, the generation of superchiral fields has mainly relied on chiral plasmonic resonators or chiral 3D architectures.[11-27] In these approaches, the difficulty lies in the external control of the sign of the chirality density which is set once for all by the shape of the resonators. Recently, other interesting approaches using achiral plasmonic[8, 11, 28-30] or dielectric[9, 31-34] resonators featuring chiral near-fields yielding some signatures of CD have been proposed. But while achiral resonators hold the exciting property to control the average sign of C by either changing the angle of illumination[35] or polarization[36-39] of the incoming light, this effect is obtained at the cost of a non-uniform distribution of the chirality density around the photonic nanostructure, diminishing the overall impact and reducing the experimental interest of these types of achiral nano-antennas. Here, we theoretically and numerically propose to generate a pure and nanoscale hot spot of chiral light in the near-field by using a single achiral plasmonic nano-resonator behaving, in agreement with Babinet principle,[40-43] as a magnetic dipole and based on a rectangular nanoslit (NS).[44-46] When observed in transmission, this hot-spot of chiral light, generated by a linearly polarized excitation, is free of any background from the excitation beam and its sign can be tuned by either changing the wavelength or the polarization angle of the incoming light. Moreover, we demonstrate the distinct advantage of using a NS over other achiral plasmonic antennas used so far and based on nano-rod (NR) geometries, in terms of yielding uniform superchiral light. Finally, to fully understand the physical mechanisms underlying these unique properties, we demonstrate, via an analytical model, that these effects are due to the magnetic dipolar nature of our NS which constructively interferes with the incident optical excitation. The resulting nanovolume of superchiral light can then serve as a nanosource to excite chiral molecules in solution present only in the nanoslit. Therefore, detecting the fluorescence or autofluorescence of the molecules present in this chiral hotspot, particularly the difference in emission between a right chiral and a left chiral excitation, makes it possible to measure the CD for a small number of molecules. Indeed the sensitivity of detectors today allows them to detect single photons and plasmonic nano-apertures are particularly well adapted to detect single molecules in solution.[47] Thus, the nanoplatform we propose here to generate a nanosource of tunable superchiral light, free of background noise, provides a roadmap for optimizing signal-to-noise ratios in emission-based rather than absorption-based circular dichroism measurements to achieve single molecule sensitivity.

# RESULTS AND DISCUSSION

The rectangular NS used in this study is schematically represented in **Figure 1**a (see Supporting Information SI.11 for simulation details). It is made in a thin gold layer of thickness H=40 nm deposited on a glass substrate. The width W of the NS is fixed at 20 nm and its length L varies from 50 to 90 nm in steps of 10 nm. The excitation is performed from the glass substrate by a linearly polarized plane wave, with a polarization angle θ, and propagates along the positive Z axis (**Figure 1**a). θ was chosen with respect to the short axis of the NS, light polarized at θ = 0° being along the X axis of the NS (**Figure 1**b). In this paper, we mainly observe the distribution of electromagnetic fields in a plane within the rectangular NS (XY plane when Z=20 nm) and for two points, point A is the geometric center (0,0) of the NS and point B is 5 nm away from the short edge and on the y axis. These two points correspond to the positions of the maximum electric and magnetic fields in the NS, respectively. In the FDTD calculations, the mesh size is set at 1 nm to ensure an accurate optical response of the NS. **Figures 1**c and **1**d show the spectral responses of the NS for different lengths L. These spectra represent the enhancement of the electric (**Figure 1**c) and magnetic (**Figure 1**d) fields normalized to the incident wave, in the center (point A) and side (point B) of the NS, 20 nm above the glass substrate and for θ = 45°, respectively. We observe resonant behaviours of the electric and magnetic fields with a linear red shift when the length of the NS increases. In the following, we will focus on an NS length of 80 nm, yielding a resonance at 680 nm.

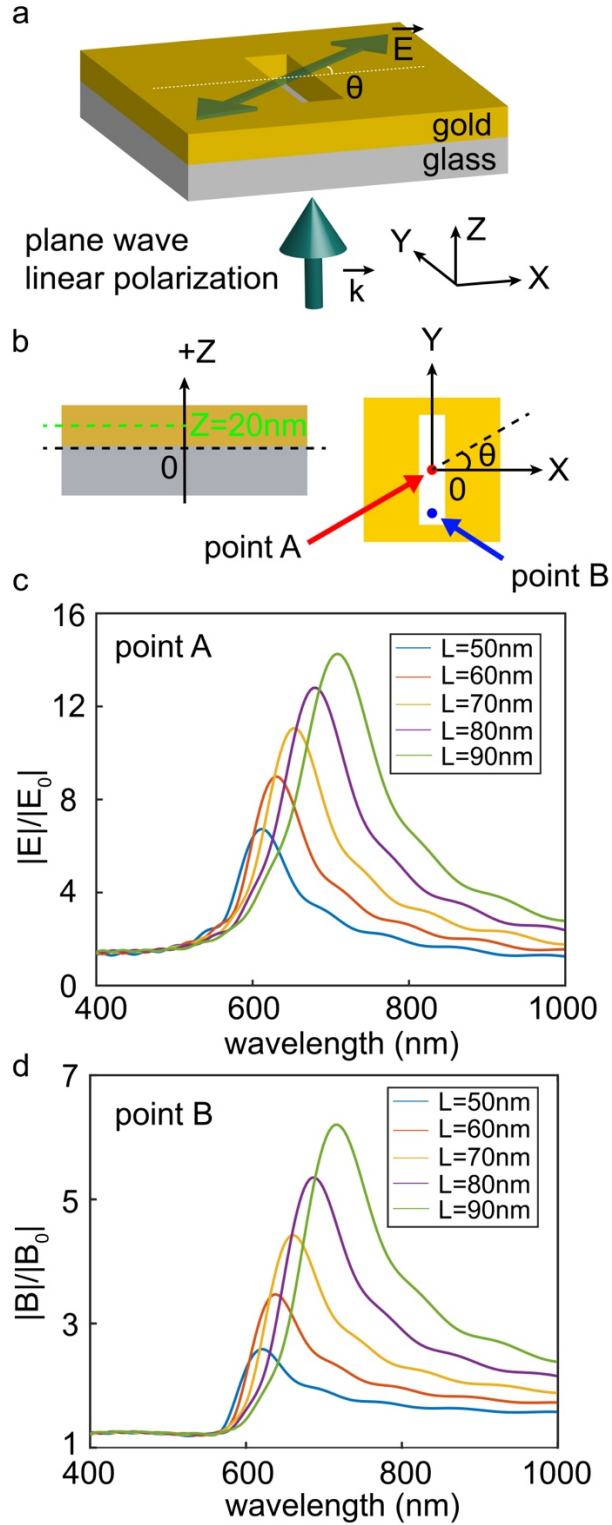

**Figure 1**. *Schematic representations and spectral responses of the magnetic dipolar NS. a) 3D and b) 2D representation of the rectangular NS in a thin gold layer of 40 nm. The vector **k** represents the direction of propagation of the linearly polarized incident*

*plane wave, and θ, the angle of this polarization with respect to the NS's transverse axis (Ox). c) Electric and d) magnetic spectral responses for different lengths of the optical NS shown in a) and b), for a width of 20 nm and θ = 45°. The spectra represent the electric and magnetic fields normalized by the incident wave, respectively, at points A (geometric center of the NS in the three dimensions of space) and B (at the center of the NS in XZ and 5 nm from the edge of the NS in Y), as shown in the inset in b).*

In order to describe the electromagnetic behavior of this NS further, **Figures 2**a and **2**b provide the spatial distributions of the electric and magnetic field enhancements in the observation plane of the NS, and excited for θ=45° at the wavelength λ=680 nm. We see, in good agreement with Babinet's principle, that the distribution of the electric and magnetic fields are opposite with respect to an electric dipole antenna such as a plasmonic NR[42, 43]. In particular, the **E**-field is concentrated homogeneously in the center of the NS while the **B**-field is maximum at the edges. Also, these electric and magnetic fields are increased by a factor of about 13 and 6 with respect to the incoming light, respectively. Furthermore, **Figures 2**c and **2**d display the time-average electric and magnetic energy density $U_e = \frac{\varepsilon_0}{4}|E|^2$ and $U_b = \frac{|B|^2}{2\mu_0}$, where $\mu_0$ is the permeability of vacuum, at the points A and point B of the NS as a function of the wavelength of the incoming plane wave and of the angle of its linear polarization with respect to the NS. We see that $U_e$ and $U_b$ are maximum at λ = 680 nm for a polarization angle of θ = 0° corresponding to a polarization along the short axis of the NS, in good agreement with experiments demonstrating higher transmission for that polarization[44, 46]. These observations support the hypothesis of a magnetic dipole character for the NS[45, 46]. It is important to notice that the electric and magnetic energy densities are the same inside the NS for opposite angles ±θ.

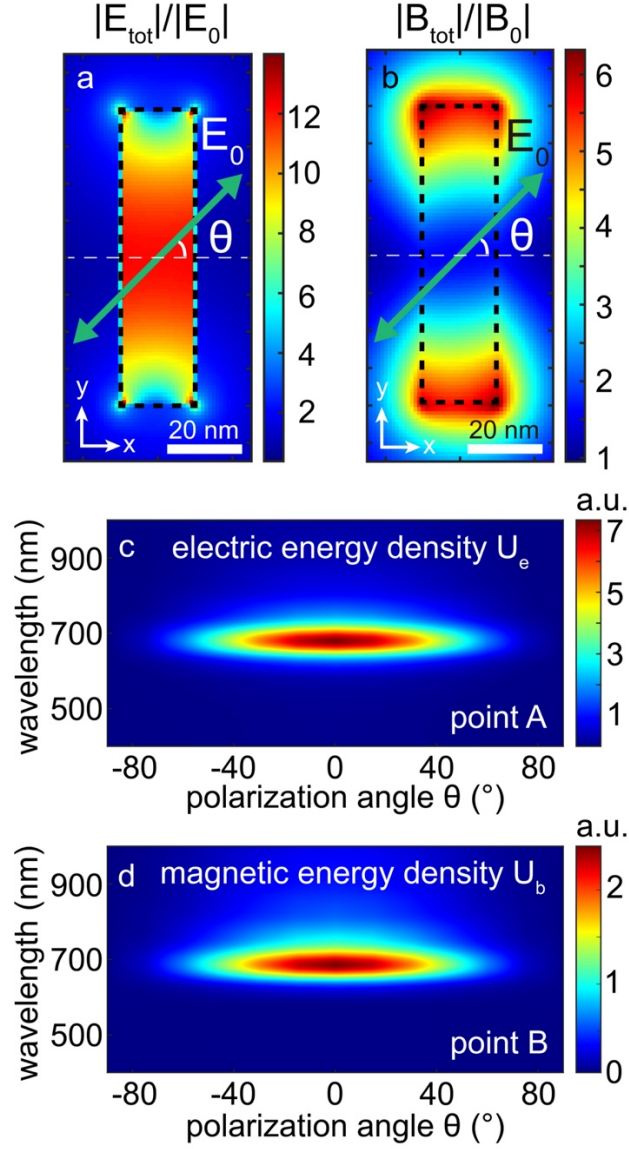

**Figure 2**. *Electromagnetic field distribution and spectral energy responses. Spatial distribution of a) electric and b) magnetic fields in an XY plane at the center of the NS in Z, at lambda=680 nm, and for a polarization angle θ = 45°. The dashed lines represent the edges of the NS. Spectral responses of c) the electric and d) the magnetic energy densities as a function of the incident polarization angle θ and wavelength respectively at points A and B (inset of **Figure 1**b).*

From the electromagnetic fields calculated and displayed in **Figures 2**a and **2**b, we have computed the chirality density enhancement $C_{enh}$ in the NS, defined as $C/|C_{cpl}|$, with $C_{cpl}$ the chirality density of a circularly polarized light, without the aperture, carrying the same power as the linearly polarized excitation impinging on the NS. **Figure 3**a presents $C_{enh}$ at the geometric center point A of the NS as a function of the wavelength for two polarization angles (45° and -45°). The results are striking. Indeed, as we can see, for opposite polarization angles θ=±45°, the sign of chirality density is reversed, something that is true for any pairs of opposite angle θ but with a lower magnitude of $C_{enh}$ (see Supporting Information SI.1). Also, and surprisingly, the sign of the chirality density is changed when crossing the resonant frequency of 680 nm, $C_{enh}$ being close to 0 at the resonance. Two opposite extrema are then observed on each side of the resonance, one at 645 nm and the other at 718 nm, with chirality density enhancements of respectively 4.5 and 3.5, in absolute value. From these results, the spatial distributions of the chirality density in the observation plane and for the wavelengths of 645, 680 and 718 nm are shown in **Figure 3**b and **3**c, for θ = +45° and θ = -45°, respectively. Remarkably, while the local chirality density distributions around plasmonic antennas have always been observed as non-uniform[36, 38, 39], in this case, they are perfectly homogeneous at the wavelengths of the two extrema (645 and 718 nm). Also, as we can see in **Figures 3**b and **3**c, the sign of the chirality density changes altogether by switching the polarization from +45° to -45°. Therefore, the NS provides a pure nanosource of chiral light tunable using either the incoming wavelength or polarization and free of any background (see **Figure SI2** for transmission enhancement), something that no other system can currently achieve. Furthermore, the chirality distribution averages 0 within the NS at the resonance wavelength for a linearly polarized excitation. However, it reaches a maximum at resonance for a circularly polarized excitation, and the sign of C can be changed depending on the helicity of the light (see **Figure SI3**). Therefore, depending on the excitation polarization of this NS, the chirality density can be fully controlled and enhanced at the nanoscale over a wide spectral range. Indeed, when measuring circular dichroism experimentally, photoelastic modulators are used to modulate the incident light polarization on chiral molecules, and they produce all possible polarizations, from linear (+45 and -45°) to right and left circular. Therefore, by demodulating the signal, the NS gives access to a very large spectral range covering about 300 nm, centered on the resonance wavelength, and whose sign of the chirality density can be fully controlled over this range. Moreover, by changing the NS's length or material, the whole visible spectrum can be covered, as well as the ultraviolet where molecules efficiently absorbe light[48] (see **Figure SI4**). Also, fabrication defects of the NS, such as rounded corners, would only marginally affect the optical and chiral response of the nanostructure (see **Figure SI5**). Finally, and remarkably, placing this type of plasmonic nanostructure in a lattice does not change the chirality density distribution in the NS nor its tunability, thus facilitating the use of NS experimentally (See **Figure SI**7). Note that defects in the nanofabrication or granularity of the gold layer could lead to an intrinsic chirality signal. One way to avoid this would be to use single crystal gold flakes and to nanofabricate these antennas by high aspect ratio lithography techniques such as a Helium focused ion beam. Also, using these antennas in an array would allow the removal of individual defects from one nanoslit to another. Finally, a characterization of the CD of these magnetic antennas without the presence of chiral molecules would allow calibrating of the blank behavior of these antennas.

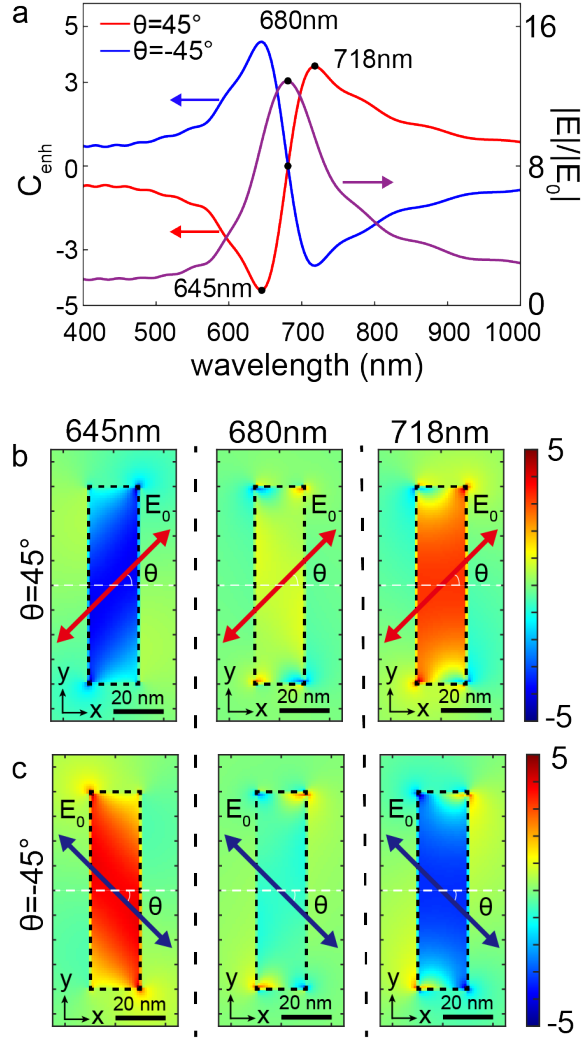

**Figure 3**. *Spectral and spatial responses of the chirality density. a) Spectral responses in chirality density (blue and red lines) and intensity (purple line) for a NS of length 80 nm, width 20 nm, for two opposite angles θ = ±45° and at point A (inset of **Figure 1**b). Spatial distribution of the chirality density in an XY plane within the NS in Z, for three characteristic wavelengths and for b) θ = +45° and c) θ = -45°.*

To further investigate these findings, and to compare the results of our magnetic dipole antenna with its electric counterpart, the volumetric chirality density $C_{vol}$ generated by the NS is compared to that of a gold NR, having the dimensions W=20 nm, H=40 nm and L=70 nm, and creating a $C_{enh}$ maximum at $\lambda$ = 645 nm (see Supporting Information SI.6). For this purpose, we simulated the electromagnetic response of both types of antennas (NS and NR) supported by glass and excited by a plane wave propagating along the positive Z-axis from the substrate. The wavelength was set to 645 nm and the incident field was linearly polarized with an angle θ with respect to the short axis of the NS and NR (**Figures 4**a and **4**b). For this study, we consider how such antennas could be exploited experimentally to study chiral molecules. Thus, in the case of the NS, only the internal volume of the NS is considered useful in terms of near-field light-matter interactions, and this volume is equal to 64000 nm$^3$. On the other hand, only the part surrounding the NR would be accessible to potential chiral molecules, which, for a similar volume, corresponds to a thickness, of 6 nm around the NR (64704 nm$^3$, when neglecting the plane on which the NR sits on the glass substrate). **Figure 4**c plots $C_{vol}$ for the NS and NR as a function of the angle θ. Several observations can be made. First, we see that for the same angle, the volumetric chirality densities have opposite signs. Second, in absolute value, the NS featured $C_{vol}$ values that are increased by a factor of 9 compared to the NR, for all polarization angles. This is due to the non-uniformity of the chirality density around the NR which decreases the volumetric average $C_{vol}$, in contrast to the NS which generates a pure chiral light as illustrated by the insets of **Figure 4**c. In the insets of **Figure 4**c, the chirality density distributions were calculated in middle plane of the NS and 5 nm above the NR. Finally, the comparative advantage of the NS is not limited to the increase of the chirality density in the near field but must also be highlighted with respect to the interaction volume that a far field excitation generates. Indeed, while the chiral signal provided by the NS would be free of any background signal due to the opacity of the gold layer in the case of transmission measurements, the one generated by the NR would be added to background chiral light-matter interactions occurring during the propagation of the incoming plane wave far from the antenna. This is something that is not considered here, but which must be taken into account if one wants to use such achiral antennas experimentally.

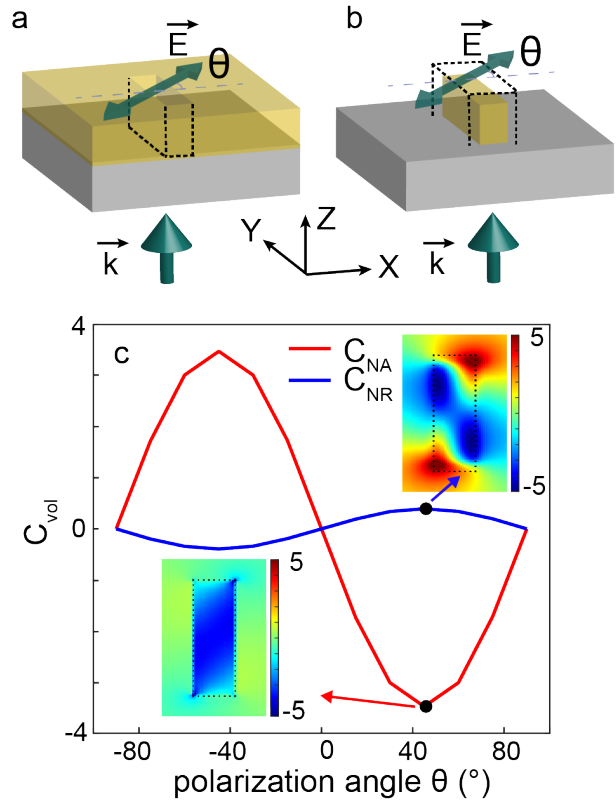

**Figure 4**. *Comparison of gold NS versus NR. Schematic representations of a) the NS and b) the NR, excited by a plane wave linearly polarized by an angle θ with respect to the transverse axis of the antennas (Ox). The dotted lines determine the volume considered in the calculation of $C_{vol.}$ c)Volumetric chirality density inside the NS and in the near field of the NR for different angles θ of the linear polarization of the excitation light. In the insets are represented the spatial distributions of the chirality density around these two antennas for θ = +45°, demonstrating the inhomogeneity of C in the case of the plasmonic NR.*

## Point-like dipole model

The wavelength and polarization tunability as well as the nanoscale homogeneity of the chirality density that our NS allows is very different from what was observed in the past in chiral[49] or achiral plasmonic nanostructures[50] such as for plasmonic NRs[36, 38, 39, 51]. The unusual properties demonstrated in this work are intrinsically linked to the magnetic dipolar character of the NS[46]. In particular, the interaction between the incident light and the magnetic dipole emission of the antenna is at the origin of these unique characteristics. In order to highlight this coupling, we use a point-like dipole model, similarly to what was carried out for the study of electric dipolar antennas [36].

The total fields radiated by a point-like magnetic dipole with moment **m**, containing both near-field and far-field contributions and decomposed in a magnetic field **B**$_{rad}$ and an electric field **E**$_{rad}$ contributions observed in the direction **n** at a distance r from the position of the dipole, are given by:[40]

$$\boldsymbol{B}_{rad} = \frac{1}{4\pi\mu_o} \left\{ k^2(\boldsymbol{n} \times \boldsymbol{m}) \times \boldsymbol{n} \frac{e^{ikr}}{r} + [3\boldsymbol{n}(\boldsymbol{n}.\boldsymbol{m}) - \boldsymbol{m}] \left( \frac{1}{r^3} - \frac{ik}{r^2} \right) e^{ikr} \right\} \quad (2a)$$

$$\boldsymbol{E}_{rad} = -\frac{Z_o}{4\pi} k^2 (\boldsymbol{n} \times \boldsymbol{m}) \frac{e^{ikr}}{r} \left( 1 - \frac{1}{ikr} \right) \quad (2b)$$

with k the wavevector, $Z_o$ the impedance of vacuum and $\mu_o$ the permeability of vacuum. The amplitude of the magnetic moment is given by $\boldsymbol{m} = \alpha^{mm} \boldsymbol{H}_o$ with $\alpha^{mm}$ the magnetic polarizability tensor and **H**$_o$ the incident magnetic field (see Supporting Information SI3). Interestingly, we see that **E**$_{rad}$ and **B**$_{rad}$ are always orthogonal; therefore, the chirality density $C_{rad} = -\epsilon_o\omega\text{Im}(\boldsymbol{E}_{rad}^* . \boldsymbol{B}_{rad})/2$ is always equal to zero in the near-field. However, the magnetic field contains a longitudinal component in the near-field that generates an elliptically polarized component that is characterized by a transverse spin angular momentum non vanishing in the near-field (non-zero spin density), as already pointed out for electric point-like dipoles.[52] This is remarkable, since in the far field, the chirality density and spin density are related and linked the same quantity which is the helicity of the propagating light. But, in the near field, due to the evanescent character of the electric and magnetic fields these two quantities are no longer proportional.[5] For a point-like dipole to generate a non-vanishing chirality density, it is required to take into account the contribution of the incident $E_o$ and $B_o$ fields as well as the radiated $E_{rad}$ and $B_{rad}$ fields so that the total fields are $E_{tot}=E_{rad}+E_o$ and $B_{tot}=B_{rad}+B_o$. The interferences between the incident and radiated fields, in amplitude, polarization, and phase, therefore allow the generation of a non-zero chirality density. Here, the resonance of the magnetic dipole moment is described with a lorentzian spectral dependence, with a phase shift that reaches π/2 when the dipole is driven at resonance. The magnetic dipole, oriented along x, is then excited by the magnetic field of a linearly polarized plane wave, propagating along z and polarized at 45° with respect to the orientation of the magnetic dipole. Taking into account that in the near field k.r << 1, the expression of the chirality density can then be simplified to the first order in k.r. Considering that in the case of the NS the field enhancement ξ =$E_{tot}$/$E_o$=13 (**Figures 1**c and 2a), the expression of $C_{tot}$ becomes $C_{tot} = -\varepsilon_o\mu_o\omega Im\left(\left(E_0 + (\xi - 1)E_{rad}\right)^* . \left(B_0 + B_{rad}\right)\right)/2$ (derivation detailed in SI 3) which yields :

$$C_{tot}(\pm 45°) = \pm \frac{\omega}{8\pi Z_o c^2 r^3}(\xi k . r . \alpha' - \alpha'') + 0(k . r^2) \qquad (2)$$

where $\alpha'$ and $\alpha''$ are the real and imaginary parts of the magnetic polarizability, respectively.

**Figure 5** shows the spectral dependence of $C_{tot}$ derived from equation 2, 20 nm above the dipole in z and for both polarization angles +/-45°. The results observed are in excellent agreement with the spectral responses of the NS (**Figure 3**a). An inversion of the sign of the chirality density is found at the resonance of the magnetic dipole, with two spectral extrema on each side of the resonance's wavelength. Also, the sign of the chirality density changes with the angle of the polarization of the plane wave incident on the dipole. These observations are perfectly explained by Equation (2). In particular, the spectral dependence is directly related to the large field enhancement factor which allows the real part of the magnetic dipole moment to dominate the spectral variations of $C_{tot}$. One can expect the spectral variations of $C_{tot}$ to be dominated by the imaginary part of the magnetic dipole moment for smaller electric field enhancements (see Supporting Information SI.3).

Using the expression of $C_{tot}$ (Equation (2)), the chirality density distributions in an xy plane 20 nm away from the dipole were calculated and are shown in **Figures 5**b and **5**c. We used an incident polarization of 45° and -45°, a field enhancement $\xi = 13$ and wavelengths of 645 nm, 680 nm and 718 nm.

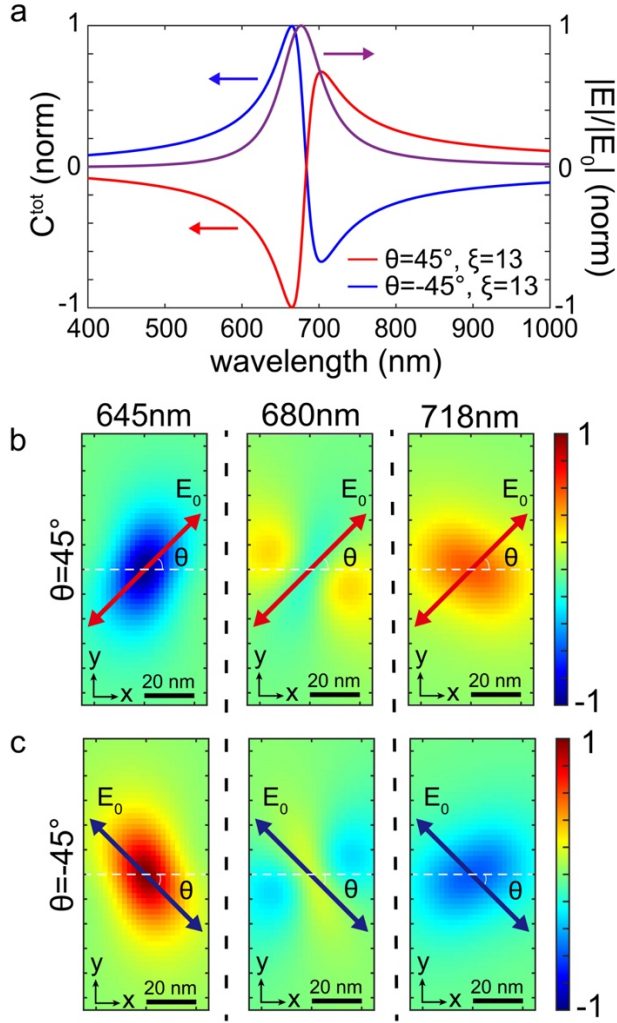

**Figure 5.** *Spectral and spatial chirality density responses obtained using a magnetic dipole model. (a) Spectral chirality density and intensity responses 20 nm above a magnetic dipole interacting with a linearly polarized plane wave propagating along the positive Z axis and having a polarization angle relative to the dipole of θ = ±45°. Spatial distribution of the chirality density in an XY plane, 20 nm above the magnetic dipole, for the three characteristic wavelengths corresponding to the extrema and the resonance and for (b) θ = +45° and (c) θ = -45°.*

Once again, we see a perfect agreement between the numerical results obtained for the NS simulations (**Figures 3**b and **3**c) and for the analytical approach using a dipole model (**Figures 5**b and **5**c). In particular, a uniform chirality density distribution is obtained for each extrema ($\lambda$ = 645 nm and 718 nm). At resonance ($\lambda$ = 680 nm), the chirality density is no longer uniform, much smaller and vanishes on the average exactly like for the numerical results of the NS at resonance. These observations are very different from those made for NRs where the field enhancement was not taken into account.[36, 38, 39] These observations offer a new roadmap in the design of nanostructures featuring enhanced chirality densities controlled externally by linearly polarized light and wavelengths, in particular through the optimization of magnetic dipolar resonances.

## CONCLUSIONS

In conclusion, we have demonstrated numerically and theoretically that the coupling between an achiral linearly polarized propagating wave and a magnetic dipolar achiral photonic NS produces a nanosource of purely superchiral light free of any background. Furthermore, we have demonstrated that the sign of the created chirality density was externally tunable simply by changing the excitation wavelength or the angle of the linear polarization incident on the structure. Finally, we have demonstrated that the useful chirality density produced by this photonic NS exceeded by one order of magnitude that of other types of achiral nanotructures such as gold NRs behaving as electric dipoles, highlighting the contribution of the magnetic dipole in the case of the nanoslit. By creating externally and on-demand tunable and background-free superchiral light nanosources, this work opens unique perspectives for the ultrasensitive detection of chiral molecules in solution at the nanoscale, with exciting perspectives to reach single-molecule sensitivity by monitoring, for instance, chiral luminescence instead of absorption.

**Supporting Information**

The Supporting Information is available free of charge

Additional information about the increase of chirality density in a nanoslit (NS) as a function of angle and wavelength, the transmission enhancement through the NS, the chirality density under circular polarization excitation, the spectral responses in volumetric chirality density for NSes of different lengths made in a gold layer and aluminum, the spectral responses in volumetric chirality density for two types of NS, one with 90° angled corners and the other with rounded corners, the chirality density enhancement study for a nanorod, the spectral response in volumetric chirality density for an NSes array, the derivation of the point-like dipole model, the simulation parameters.

**Acknowledgements**

We acknowledge the financial support from the Agence national de la Recherche (ANR-20-CE09-0031-01), from the Institut de Physique du CNRS (Tremplin@INP 2020) two of the authors (L.C. and X.Y.) acknowledge the support of the China Scholarship Council.

**Conflict of Interest**

The authors declare no conflict of interest.

**Table of content graphic**

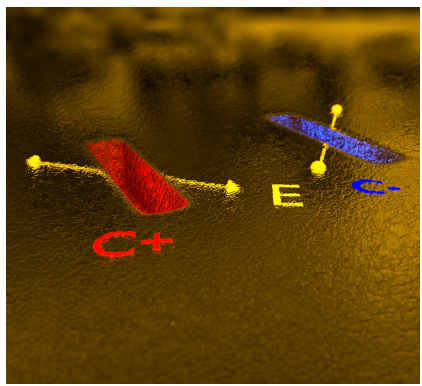

# References

1. Laska, M.; Teubner, P. *Chemical senses* **1999,** 24, (2), 161-170.
2. Davies, N. M.; Teng, X. W. *Advances in Pharmacy* **2003,** 1, (3), 242-252.
3. Tang, Y.; Cohen, A. E. *Phys. Rev. Lett.* **2010,** 104, (16), 163901.
4. Lipkin, D. M. Journal of Mathematical Physics **1964,** 5, (5), 696-700.
5. Bliokh, K. Y.; Nori, F. *Phys. Rev. A* **2011,** 83, (2), 021803.
6. Coles, M. M.; Andrews, D. L. *Phys. Rev. A* **2012,** 85, (6), 063810.
7. Poulikakos, L. V.; Gutsche, P.; McPeak, K. M.; Burger, S.; Niegemann, J.; Hafner, C.; Norris, D. J. *ACS photonics* **2016,** 3, (9), 1619-1625.
8. Lee, S.; Yoo, S.; Park, Q.-H. *ACS Photonics* **2017,** 4, (8), 2047-2052.
9. Raziman, T.; Godiksen, R. H.; Muller, M. A.; Curto, A. G. *ACS Photonics* **2019,** 6, (10), 2583-2589.
10. Both, S.; Schäferling, M.; Sterl, F.; Muljarov, E. A.; Giessen, H.; Weiss, T. *ACS Nano* **2022,** 16, (2), 2822-2832.
11. Govorov, A. O.; Fan, Z.; Hernandez, P.; Slocik, J. M.; Naik, R. R. *Nano. Lett.* **2010,** 10, (4), 1374-1382.
12. Abdulrahman, N. A.; Fan, Z.; Tonooka, T.; Kelly, S. M.; Gadegaard, N.; Hendry, E.; Govorov, A. O.; Kadodwala, M. *Nano. Lett.* **2012,** 12, (2), 977-983.
13. Hendry, E.; Mikhaylovskiy, R.; Barron, L.; Kadodwala, M.; Davis, T. *Nano. Lett.* **2012,** 12, (7), 3640-3644.
14. Kaschke, J.; Gansel, J. K.; Wegener, M. *Opt. Express* **2012,** 20, (23), 26012-26020.
15. Droulias, S.; Yannopapas, V. *J. Phys. Chem. C* **2013,** 117, (2), 1130-1135.
16. Yin, X.; Schäferling, M.; Metzger, B.; Giessen, H. *Nano. Lett.* **2013,** 13, (12), 6238-6243.
17. Zhukovsky, S. V.; Chigrin, D. N.; Kremers, C.; Lavrinenko, A. V. *Photonics and Nanostructures-Fundamentals and Applications* **2013,** 11, (4), 353-361.
18. Kondratov, A.; Gorkunov, M.; Darinskii, A.; Gainutdinov, R.; Rogov, O.; Ezhov, A.; Artemov, V. *Phys. Rev. B* **2016,** 93, (19), 195418.
19. Schäferling, M.; Engheta, N.; Giessen, H.; Weiss, T. *ACS photonics* **2016,** 3, (6), 1076-1084.
20. Špačková, B.; Wrobel, P.; Bocková, M.; Homola, J. *Proceedings of the IEEE* **2016,** 104, (12), 2380-2408.
21. Wang, Z.; Wang, Y.; Adamo, G.; Teh, B. H.; Wu, Q. Y. S.; Teng, J.; Sun, H. *Advanced Optical Materials* **2016,** 4, (6), 883-888.
22. García-Guirado, J.; Svedendahl, M.; Puigdollers, J.; Quidant, R. *Nano. Lett.* **2018,** 18, (10), 6279-6285.
23. Ai, B.; Luong, H. M.; Zhao, Y. *Nanoscale* **2020,** 12, (4), 2479-2491.
24. Luong, H. M.; Pham, M. T.; Nguyen, T. D.; Zhao, Y. *J. Phys. Chem. C* **2020,** 125, (1), 716-723.
25. Petronijevic, E.; Ali, H.; Zaric, N.; Belardini, A.; Leahu, G.; Cesca, T.; Mattei, G.; Andreani, L.; Sibilia, C. *Optical and Quantum Electronics* **2020,** 52, (3), 1-10.
26. Qu, Y.; Bai, Y.; Aba, T.; Ullah, H.; Abudukelimu, A.; Huang, J.; Gou, T.; Li, J.; Zhang, Z. *J. Phys. Chem. C* **2020,** 124, (25), 13912-13919.

27. Wu, B.; Wang, M.; Sun, Y.; Wu, F.; Shi, Z.; Wu, X. *Advanced Composites and Hybrid Materials* **2022,** 5, (3), 2527-2535.
28. Slocik, J. M.; Govorov, A. O.; Naik, R. R. *Nano. Lett.* **2011,** 11, (2), 701-705.
29. Maoz, B. M.; Ben Moshe, A.; Vestler, D.; Bar-Elli, O.; Markovich, G. *Nano. Lett.* **2012,** 12, (5), 2357-2361.
30. Nesterov, M. L.; Yin, X.; Schäferling, M.; Giessen, H.; Weiss, T. *Acs Photonics* **2016,** 3, (4), 578-583.
31. García-Etxarri, A.; Dionne, J. A. *Phys. Rev. B* **2013,** 87, (23), 235409.
32. Garcia-Guirado, J.; Svedendahl, M.; Puigdollers, J.; Quidant, R. *Nano. Lett.* **2019,** 20, (1), 585-591.
33. Hu, J.; Lawrence, M.; Dionne, J. A. *ACS Photonics* **2019,** 7, (1), 36-42.
34. Feis, J.; Beutel, D.; Köpfler, J.; Garcia-Santiago, X.; Rockstuhl, C.; Wegener, M.; Fernandez-Corbaton, I. *Phys. Rev. Lett.* **2020,** 124, (3), 033201.
35. Guth, N.; Gallas, B.; Rivory, J.; Grand, J.; Ourir, A.; Guida, G.; Abdeddaim, R.; Jouvaud, C.; De Rosny, J. *Phys. Rev. B* **2012,** 85, (11), 115138.
36. Schäferling, M.; Yin, X.; Giessen, H. *Opt. Express* **2012,** 20, (24), 26326-26336.
37. Yao, H.; Zhong, S.; Yi, H.; Cui, D.; Pan, C.-L. *IEEE Photonics Journal* **2017,** 9, (3), 1-9.
38. Hashiyada, S.; Narushima, T.; Okamoto, H. *ACS Photonics* **2018,** 5, (4), 1486-1492.
39. Hashiyada, S.; Narushima, T.; Okamoto, H. *ACS Photonics* **2019,** 6, (3), 677-683.
40. Jackson, J. D., Classical electrodynamics. American Association of Physics Teachers: 1999.
41. Grosjean, T.; Mivelle, M.; Baida, F.; Burr, G.; Fischer, U. *Nano. Lett.* **2011,** 11, (3), 1009-1013.
42. Ögüt, B.; Vogelgesang, R.; Sigle, W.; Talebi, N.; Koch, C. T.; van Aken, P. A. *ACS Nano* **2011,** 5, (8), 6701-6706.
43. Singh, A.; Calbris, G.; van Hulst, N. F. *Nano. Lett.* **2014,** 14, (8), 4715-4723.
44. Degiron, A.; Lezec, H.; Yamamoto, N.; Ebbesen, T. *Opt. Commun.* **2004,** 239, (1-3), 61-66.
45. Curto, A. G. Optical antennas control light emission. Universitat Politècnica de Catalunya, 2014.
46. Park, Y.; Kim, J.; Roh, Y.-G.; Park, Q.-H. *Nanophotonics* **2018,** 7, (10), 1617-1636.
47. Punj, D.; Mivelle, M.; Moparthi, S. B.; van Zanten, T. S.; Rigneault, H.; van Hulst, N. F.; García-Parajó, M. F.; Wenger, J. *Nat. Nanotechnol.* **2013**.
48. Leite, T. R.; Zschiedrich, L.; Kizilkaya, O.; McPeak, K. M. *Nano. Lett.* **2022,** 22, (18), 7343-7350.
49. Hendry, E.; Carpy, T.; Johnston, J.; Popland, M.; Mikhaylovskiy, R.; Lapthorn, A.; Kelly, S.; Barron, L.; Gadegaard, N.; Kadodwala, M. *Nat. Nanotechnol.* **2010,** 5, (11), 783-787.
50. Alizadeh, M.; Reinhard, B. r. M. *ACS Photonics* **2015,** 2, (3), 361-368.
51. Davis, T.; Hendry, E. *Phys. Rev. B* **2013,** 87, (8), 085405.
52. Neugebauer, M.; Banzer, P.; Nechayev, S. *Science advances* **2019,** 5, (6), eaav7588.

# An achiral magnetic photonic antenna as a tunable nanosource of superchiral light

Lingfei Cui[1], Xingyu Yang[1], Benoît Reynier[1], Catherine Schwob[1], Sébastien Bidault[2], Bruno Gallas[1,*] and Mathieu Mivelle[1,*]

[1]Sorbonne Université, CNRS, Institut des NanoSciences de Paris, INSP, F-75005 Paris, France

[2]Institut Langevin, ESPCI Paris, PSL University, F-75005 Paris, France

*Corresponding authors:

bruno.gallas@insp.jussieu.fr

mathieu.mivelle@sorbonne-universite.fr

ORCID: 0000-0002-0648-7134

# SUPPORTING INFORMATION

**S.I.1/ Increase of chirality density in a nanoslit (NS) as a function of angle and wavelength**

The enhancement of the chirality density $C_{enh}$ in the NS was calculated at point A in the center of the NS (**Figure SI1**a) for any polarization angle θ and the wavelength range from 400 to 1000 nm (**Figure SI**1b). It can be seen that the sign of the chirality density is reversed when the sign of the polarization angle is inverted. We can also observe that the chirality density changes sign around the resonance (680 nm). The chirality density disappears when the incident polarization is along the short NS axis (x-axis or θ=0°) or the long NS axis (y-axis or θ=90°).

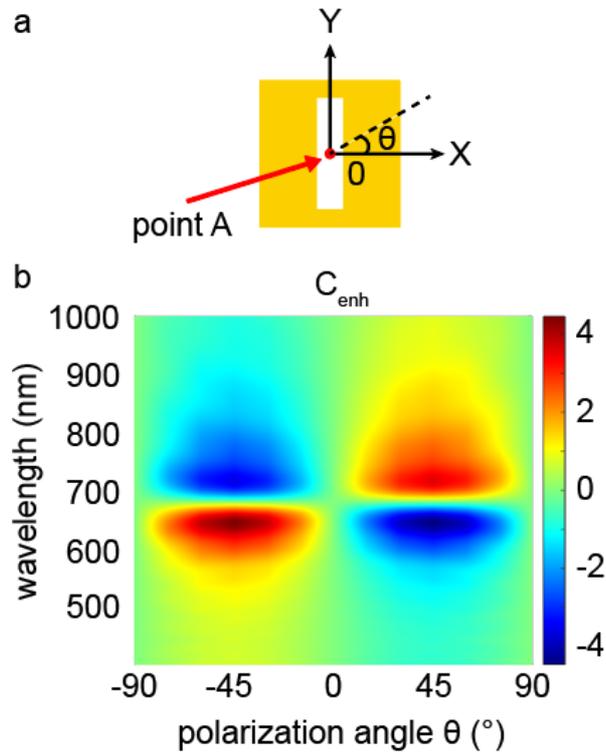

**Figure SI1.** *a) Position of point A in the NS where $C_{enh}$ was calculated. b) Variations of $C_{enh}$ as a function of the incident polarization angle θ and wavelength in the center of the NA.*

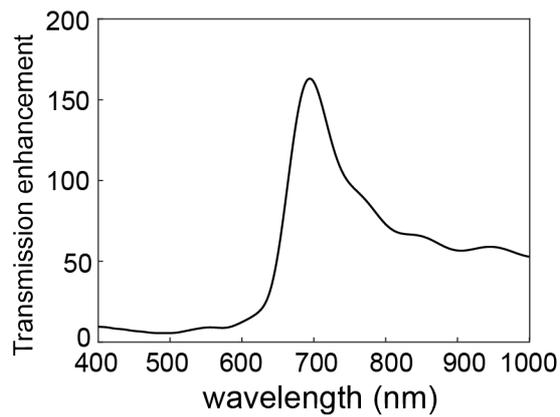

**Figure SI2.** *Transmission through the nanoslit normalized by the signal transmitted through the same surface made of a uniform gold layer.*

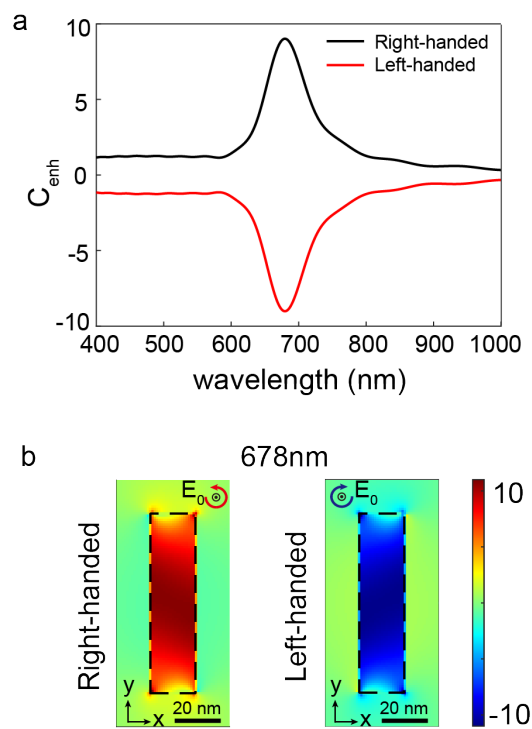

**Figure SI3**. *a) Spectral responses in volumetric chirality density for excitation of the NS by a right circular polarization (black curve) or left circular polarization (red curve). b) Spatial distribution of the chirality density at the center and at the resonance of the NS for these two excitation polarizations.*

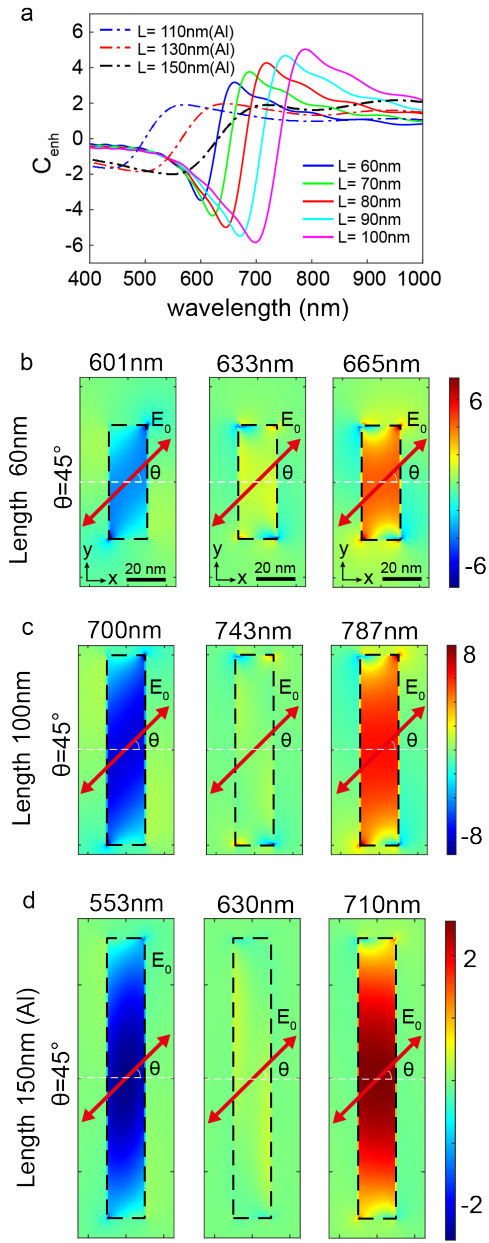

**Figure SI4**. *a) Spectral responses in volumetric chirality density for NSes of different lengths made in a gold layer (solid curves) and aluminum (dashed curves). b) Spatial distribution of chirality density at the center of several representative NSes for linear polarization at 45° from the short axis of the NSes and wavelengths corresponding to different extrema of a).*

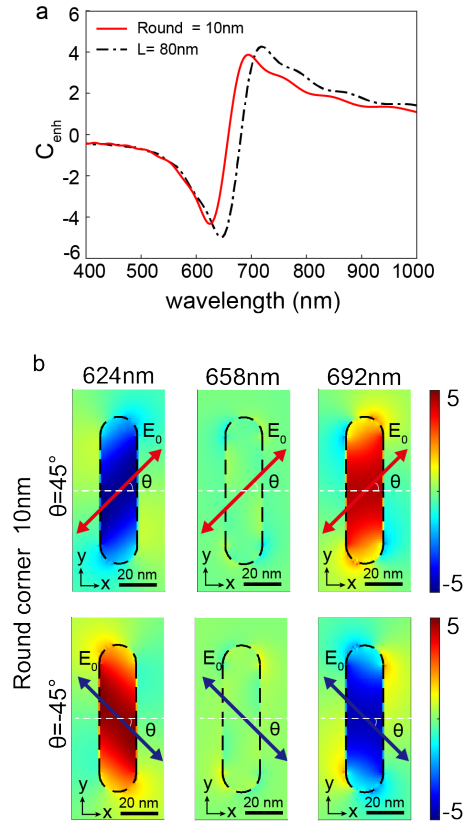

**Figure SI5**. *a) Spectral responses in volumetric chirality density for two types of NS, one with 90° angled corners and the other with rounded corners. b) Spatial distribution of the chirality density at the center of the NS with rounded corners for different polarization angles and wavelengths corresponding to different extrema of a).*

**S.I.6/ Chirality density enhancement study for nanorod (NR)**

We numerically calculated the spectral variations of $C_{enh}$ for an incident polarization angle θ of 45° in the wavelength range 400-1000 nm for an NR consisting of gold deposited on glass. $C_{enh}$ values were monitored at point C located 5 nm from one end of the NR and 20 nm above the glass substrate, where the increase in the electric field was most significant (**Figure SI**6a). We first determined the dimensions of an NR that would have a $C_{enh}$ extremum at the same wavelength as that observed at the lowest NS wavelengths, i.e., at 645 nm (Figure 3a). For this purpose, we used a width (W) of 20 nm and a height (H) of 40 nm and adjusted the length (L). It was found that $C_{enh}$ was maximum at 645nm for a length L=70 nm (**Figure SI**6b). The determination of $C_{vol}$ shown in Figure 4c was obtained with these dimensions. In contrast to NS (**Figure SI**1b), it can be seen that the sign of the chirality density is always the same throughout the spectral range for a given polarization angle (**Figure SI**6b). As for NA, the sign is reversed when the polarization angle changes from 45° to -45°.

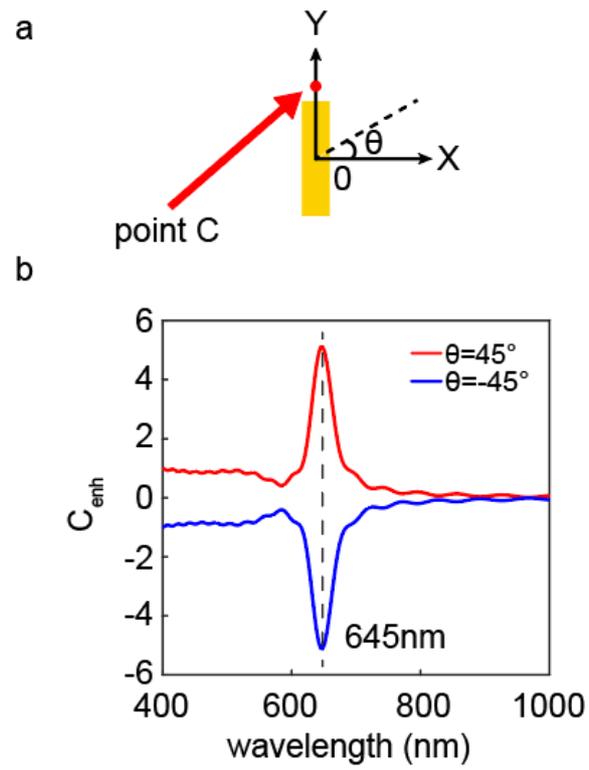

**Figure SI6**. *a) Schematic of the NR and position of the point C. b) Variations of $C_{enh}$ as a function of the wavelength for an NR with dimensions W=20nm, H=40nm and L=70nm, and θ = +/-45°.*

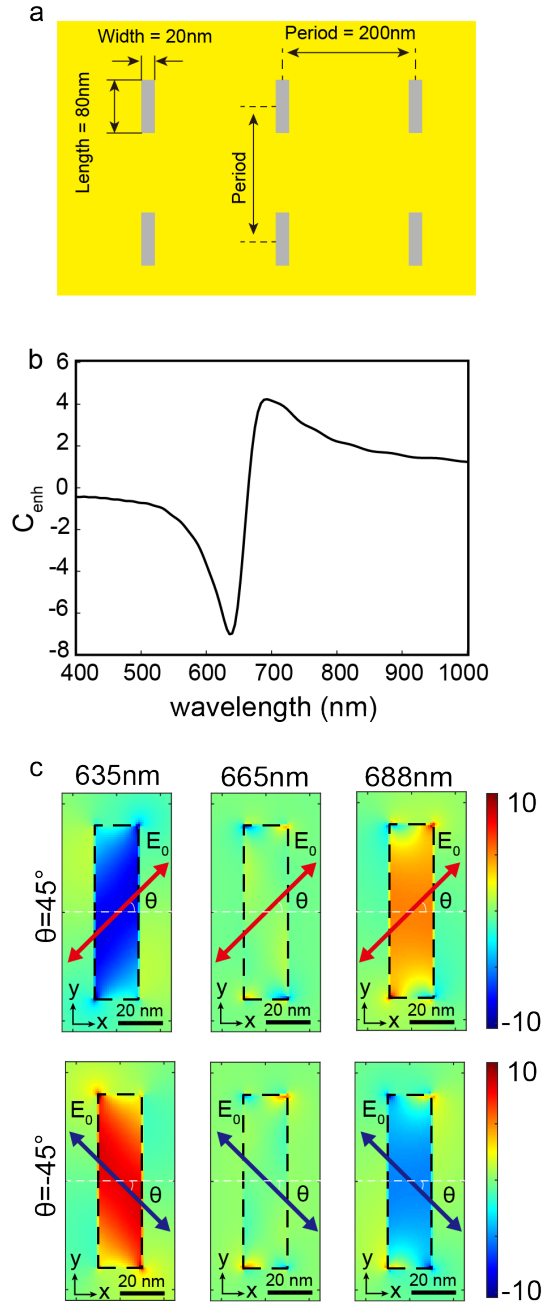

**Figure SI7**. *a) Schematic of an NS array whose period has been optimized to avoid Rayleigh anomaly problems. b) Spectral response in volumetric chirality density for one of the antennas of this array. c) Spatial distribution of the chirality density at the center of an antenna of this array for different polarization angles and wavelengths corresponding to different extrema of b).*

**S.I.8/ Derivation of the point-like dipole model**

To describe the response of the NS, we used a 3x3 diagonal magnetic polarizability tensor $\alpha^{mm}$. In our Cartesian reference frame, with the NS in the (x,y) plane having its minor axis along x, only the first element $\alpha^{mm}_{xx} = \alpha' + i\alpha'' \neq 0$ and we will describe its spectral dependence with a Lorentzian profile (Figure SI8). The Lorentz oscillator was centered at 677 nm with a broadening of 0.185 eV and had an amplitude of unity.

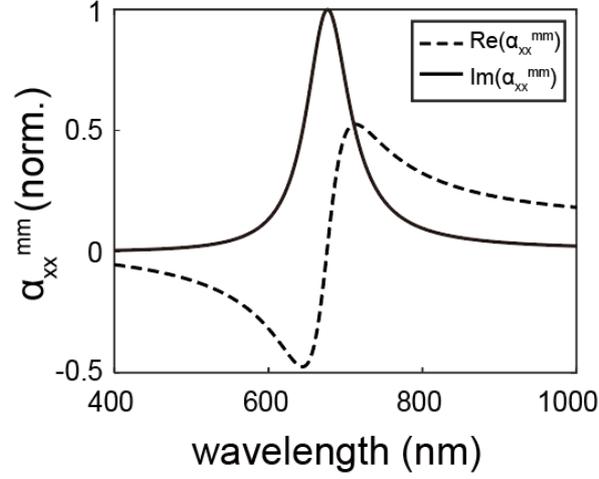

**Figure SI8**. *Spectral dependence of the real and imaginary parts of $\alpha_{xx}^{mm}$*

The magnetic moment **m** of the dipole was then obtained as :

$$\mathbf{m} = \alpha^{mm} \cdot \mathbf{H_o} \qquad (1)$$

with $\mathbf{B_o} = \mu_o \mathbf{H_o}$, $\mu_o$ being the permeability of vacuum and $\mathbf{H_o}$ the magnetic field of the incident wave. The incident plane wave propagated along the z-axis and was linearly polarized with an angle $\theta$ with respect to the x-axis. The normalized coordinates of the incident electric field were $\mathbf{E_o} = (\cos\theta, \sin\theta, 0)$. The incident magnetic field is then obtained as $\mathbf{B_o} = \frac{1}{c}\mathbf{k} \times \mathbf{E_o}$ with **k** the wavevector of the incident light.

The fields radiated by the NS were calculated using the radiation of the point-like magnetic dipole **m** (Equation (1)) which were decomposed in a magnetic field $\mathbf{B}_{rad}$ and an electric field $\mathbf{E}_{rad}$. In the direction **n** at a distance r from dipole position, the fields are given by :

$$\mathbf{B}_{rad} = \frac{1}{4\pi\mu_o}\left\{k^2(\mathbf{n} \times \mathbf{m}) \times \mathbf{n}\frac{e^{ikr}}{r} + [3\mathbf{n}(\mathbf{n} \cdot \mathbf{m}) - \mathbf{m}]\left(\frac{1}{r^3} - \frac{ik}{r^2}\right)e^{ikr}\right\} \qquad (2a)$$

$$\mathbf{E}_{rad} = -\frac{Z_o}{4\pi}k^2(\mathbf{n} \times \mathbf{m})\frac{e^{ikr}}{r}\left(1 - \frac{1}{ikr}\right) \qquad (2b)$$

with k the modulus of the wavevector of the radiated field, $Z_o$ the impedance of vacuum and c the speed of light in vacuum. These fields contain both near and far field radiation. We note that $\mathbf{E}_{rad}$ and $\mathbf{B}_{rad}$ are always orthogonal. Therefore, the chirality density $C_{rad} = -\frac{\omega}{c^2}\text{Im}(\mathbf{E}^*_{rad} \cdot \mathbf{B}_{rad})$, where ω is the angular frequency of the field, is always zero. In order to generate chirality, it is then necessary to add the contribution of the linearly polarized incident fields $(\mathbf{E_0}, \mathbf{B_0})$.

The superposition of the two linearly polarized transverse fields, incident and radiated, and the phase shift between them results in an elliptically polarized light. The phase shift reaches π/2 when the dipole is driven at resonance and it varies gradually with frequency between 0 and π due to the broadening of the Lorentzian profile chosen here for the polarizability. The total chirality density is then obtained as:[1-3]

$$C_{tot} = -\frac{\omega}{c^2} \text{Im}((\mathbf{E}_{rad}+\mathbf{E}_o)^*.(\mathbf{B}_{rad} + \mathbf{B}_o)) \qquad (3)$$

A final important ingredient of our model was to consider the enhancement of the near-fields in the NS, as suggested by our numerically calculated field maps (Figures 1c and 2a). It can be seen from these distributions that the electric field is mainly enhanced by a factor $\xi$, especially in the area where the chirality density is large, so that $\xi = \frac{\mathbf{E}_{tot}}{\mathbf{E}_o} = \frac{\mathbf{E}_{rad}+\mathbf{E}_o}{\mathbf{E}_o}$. To reproduce this effect, we applied a field enhancement factor $\xi$, to the radiated electric field only, so that $\mathbf{E}_{rad} \rightarrow (\xi - 1)\mathbf{E}_{rad}$ while $\mathbf{B}_{rad}$ remained unchanged resulting in:

$$C_{tot} = -\frac{\omega}{c^2} \text{Im}([(\xi - 1)\mathbf{E}_{rad}+\mathbf{E}_o]^*.(\mathbf{B}_{rad} + \mathbf{B}_o)) \qquad (4)$$

The total chirality density was then calculated taking all these contributions into account. Finally, using Equations (1), (2a), (2b) in (4), the total chirality density is given by:

$$C_{tot} = \frac{k}{8Z_o cr^3}[k.r.\xi(\alpha' + k.r.\alpha'') - \alpha''].\sin(2\theta) \qquad (5)$$

This relation explicitly shows the dependence of the sign of the chirality density on the polarization angle and in particular the change in sign as the polarization angle changes from $\theta$ to $-\theta$. Equation (5) also shows that the chirality density should be maximal at +/-45°. If we neglect the $k.r.\alpha_{xx,i}^{mm}$ which yields a $(k.r)^2$ term and use 45° for $\theta$, we find the expression given in the manuscript.

As presented in Figure 5, with a value of $\xi = 13$, this model reproduces quite well the variation of $C_{tot}$ as a function of wavelength and polarization angle. We present in **Figure SI**9 a the influence of the value of $\xi$ on the spectral dependence of $C_{tot}$ for a polarization angle of 45° and at 20 nm above the point-like dipole. Using a polarization angle of -45° would give exactly opposite spectral variations, they are not shown here for clarity.

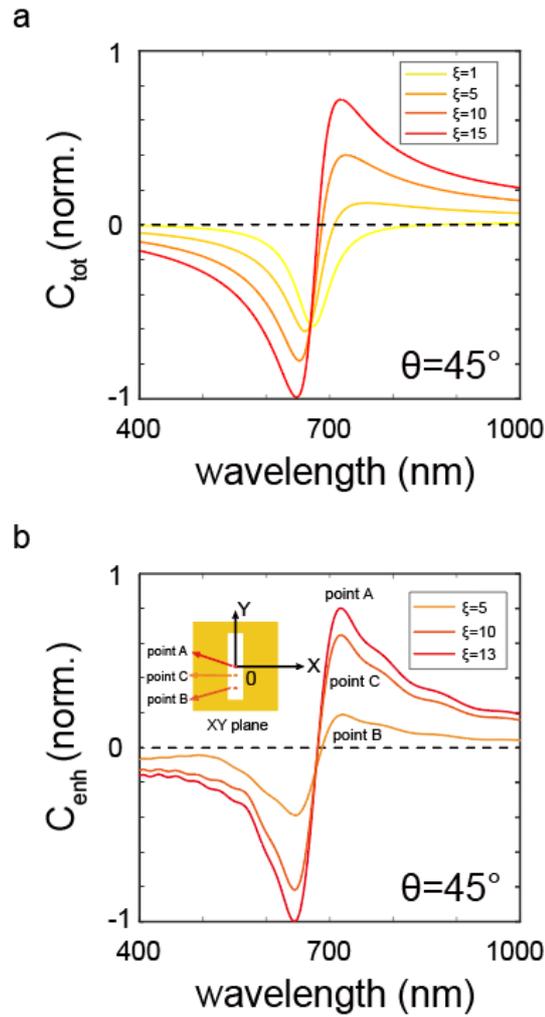

**Figure SI9**. *Spectral variation of $C_{tot}$ calculated for different values of the electric field enhancement factor $\xi$ for a) the dipolar model and b) the NS. The values of $C_{tot}$ were all normalized to the maximum value obtained (for $\xi = 15$ and wavelength = 646 nm).*

It can be seen in **figures SI9a** and **SI9b** that in the absence of electric field enhancement ($\xi$=1), $C_{tot}$ is predominantly negative with a maximum near the magnetic dipole resonance. The presence of a nonzero chirality density for $\xi$=1 originates from the magnetic field's contribution to the chirality density. As the values of $\xi$ are progressively increased, a sign change near the position of the magnetic dipole resonance is observed with values that become even larger than that observed for $\xi$=1. A comparison with the numerical calculations was tentatively made by inspecting the exact values of the electric field increase inside the NS as a function of position. Different values were found, ranging from 5 to 13. The spectral dependence of $C_{enh}$ was then plotted at these locations and is shown in **Figure SI9b**. Excellent agreement with the values found with the point dipole model was obtained, supporting the idea that electric field enhancement is indeed an important parameter in the evaluation of the chirality density near a nano-antenna.

**Figure SI10** presents the maps of $C_{tot}$ in an (x,y) plane 20 nm above the point-like dipole for a polarization angle of -45° at the extreme $C_{tot}$ values obtained at 646 nm for $\xi = 1$ and $\xi = 15$.

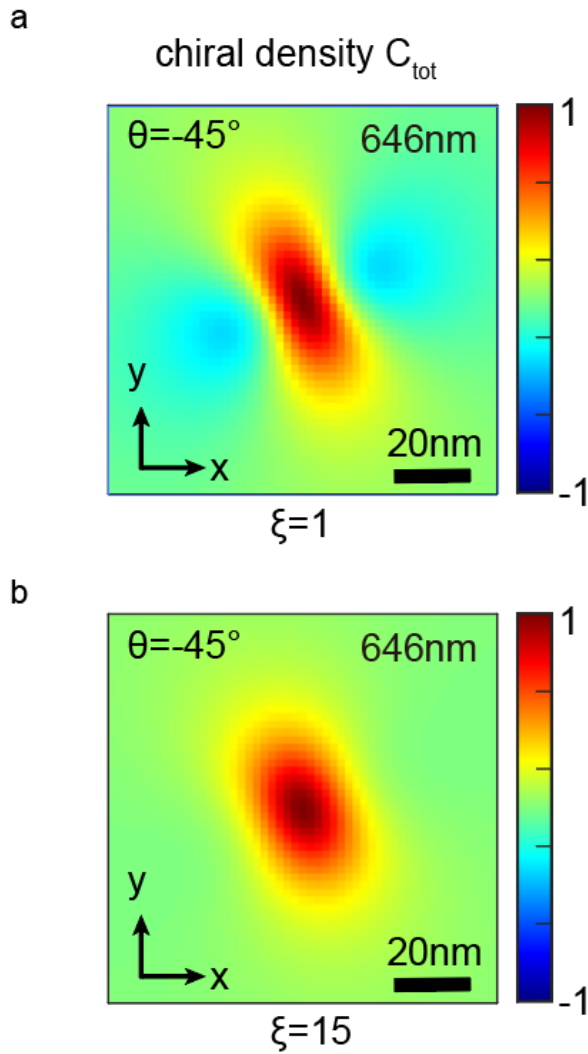

**Figure SI10**. *Spatial variations of $C_{tot}$ in a (x,y) plane 20 nm above the point-like dipole for a) $\xi = 1$ and b) $\xi = 15$.*

Upon passing resonance, the $C_{tot}$ field map for $\xi=1$ is similar to that already published in the literature for an electric dipole.[1,3] In addition to the sign change when $\xi=15$, another striking feature is that the $C_{tot}$ field distribution becomes very uniform. These field maps are in excellent qualitative agreement with those calculated numerically within the NS in Figure 3(b). This confirms that the critical parameter for obtaining spatially uniform values for the total chirality density is to generate a strong electric field enhancement as spatially extended as possible. This feature is achieved in the NS (Figure 2(a)), resulting in a uniform distribution of chirality density within the NS volume.

**S.I.11/ Simulation parameters**

The simulations were performed with the FDTD (Finite Difference Time Domain) commercial software Lumerical. They consist of a calculation window of 1 micrometer in the three directions of space. The 40 thick film of gold was placed on a semi-infinite glass substrate.

All optical constants were taken from the Lumerical database and correspond to the usual values for gold thin film [J&C] and fused silica. The source is a linearly or circularly polarized plane wave placed at -350 nm from the NS. The latter excites the nanostructure from the substrate side. This source is a pulse of 2.66 fs duration, which, via a Fourier transform, allows to excite the nano-antenna with a broad spectrum centered on the wavelength of 700 nm and a spectral width of 600 nm. The whole computational window is surrounded by a Perfectly Matching Layer (PML) to avoid unwanted reflections. The electric and magnetic optical fields were then collected in a volume of 50x110x40 nm$^3$ enclosing the NS, allowing the calculation of the chirality density in the whole nanostructure for all wavelengths contained in the optical pulse.

# References

(1) Schäferling, M.; Yin, X.; Giessen, H. Formation of chiral fields in a symmetric environment. *Opt. Express* **2012**, *20* (24), 26326-26336.

(2) Hashiyada, S.; Narushima, T.; Okamoto, H. Imaging chirality of optical fields near achiral metal nanostructures excited with linearly polarized light. *ACS Photonics* **2018**, *5* (4), 1486-1492.

(3) Hashiyada, S.; Narushima, T.; Okamoto, H. Active control of chiral optical near fields on a single metal nanorod. *ACS Photonics* **2019**, *6* (3), 677-683.